\documentclass[12pt]{article}

\usepackage{authblk}
\usepackage{amsmath}
\usepackage[margin=1in]{geometry}

\begin{document}

\title{Failed attempt to escape from the quantum pigeon conundrum}                      

\author[1,2,3]{Yakir Aharonov}
\affil[1]{\small Raymond and Beverly Sackler School of Physics and Astronomy, Tel-Aviv University, Tel-Aviv 69978, Israel}
\affil[2]{Schmid College of Science and Technology, Chapman University, Orange, California 92866, USA}
\affil[3]{Institute for Quantum Studies, Chapman University, Orange, California 92866, USA}

\author[1]{Shrobona Bagchi}

\author[4,5]{Justin Dressel}
\affil[4]{Institute for Quantum Studies, Chapman University, Orange CA 92866, USA}
\affil[5]{Schmid College of Science and Technology, Chapman University, Orange CA 92866, USA}

\author[1]{Gregory Reznik}

\author[1]{Michael Ridley}

\author[1]{Lev Vaidman}

\date{}

\maketitle

\begin{abstract}
A recent criticism by Kunstatter et al. [Phys. Lett. A  384,
  126686 (2020)] of a quantum setup violating the  pigeon counting principle [Aharonov et al. PNAS 113, 532 (2016)] is refuted. The quantum nature of the violation of the pigeonhole principle with pre- and postselection is clarified.

\end{abstract}






In a recent paper \cite{kunstatter2020escape} Kunstatter et al. analyzed the work \cite{pigeons2016} of Aharonov et al.  which  presented a violation of the pigeon counting  principle (PCP) in a particular pre- and postselected system. Kunstatter et al. claimed ``we have provided a proof that the PCP is not violated in quantum mechanics''. In this Comment we will argue that Kunstatter et al. did not resolve the PCP conundrum because they analyzed a different problem. 

The quantum pigeon conundrum is a paradox about the locations of pre- and postselected quantum particles. We prepare a particular state of three particles in two boxes and consider a possible event in which the particles are found later in another particular state. The PCP tells us that when  we  put three classical particles in  two boxes, we must find one pair in the same box. Aharonov et al. \cite{pigeons2016} argued that for   particular pre- and postselected quantum states, three quantum particles are put in two boxes, yet no two particles are in the same box. In standard quantum mechanics there is no definition of the location of a pre- and postselected particle. The proposed definitions \cite{wheeler1978past,past} led to heated controversies, and it seems that we are still very far from a consensus regarding this question. In particular, the meaning of ``no two particles present in the same box'' has to be carefully specified.   Obviously, it does not mean that the results of strong simultaneous measurements of the locations of all pairs of particles  contradict the PCP for this particular pre- and postselection, since the results of strong measurements correspond to classical elements of reality for which the PCP must hold. Elements of reality of quantum pre- and postselected systems \cite{PRLmy},  similarly to Einstein-Podolsky-Rosen elements of reality \cite{EPR} of preselected quantum systems,  cannot be all be tested simultaneously by strong measurements. What justifies assigning elements of reality to observables which cannot be measured simultaneously is that we can infer with certainty the outcome of a measurement of {\it any} of these observables. Another argument supporting assigning elements of reality to a pre- and postselected system is the robust and universal  modification of all weak (or short) interactions with the environment,  see \cite{PNAS}. We spell out  these arguments in the context of the quantum pigeon conundrum below as claims (i) and (ii). A more extensive analysis of the original proposal \cite{pigeons2016} and of  other variants of the quantum   violation of the PCP appears in \cite{pigeons_footprints}.

In their paper Kunstatter et al. first allegedly reproduced the  argument of \cite{pigeons2016} in three steps  and then showed that this three-step argument fails. Kunstatter et al. are correct that their three-step argument does not work, but are wrong that their argument faithfully reproduces the quantum pigeon conundrum as stated in Aharonov et al. \cite{pigeons2016}. The  argument of Kunstatter et al. is based on their Eq. 2. (repeated as Eq. 32 in their conclusion), but this equation, or anything equivalent to it,  does not appear in \cite{pigeons2016}. This is the essence of our Comment. We now analyze the argument of Kunstatter et al. in more detail.


Kunstatter et al. start with Eq. 1: 
\begin{align}\label{1}
 \langle +i+i+i|\Pi_{ab}|+++\rangle=0;~~~{\rm for~all}~a,b \in \{0,1,2\} ,   
\end{align}
where $\Pi_{ab}$ is a projection on the space of states in which particles $a$ and $b$ are in the same box. Equation (\ref{1})
 is indeed the basis of the argument in \cite{pigeons2016}, but   Kunstatter et al. use it differently. 
Kunstatter et al. correctly state that from (\ref{1}) follows their Eq. 2:
\begin{align}\label{2}
 |\langle +i+i+i|\Pi|+++\rangle |^2=0 ;~~~{\rm where}~
 \Pi \equiv \Pi_{01}+ \Pi_{12}+ \Pi_{02} .  
\end{align}
This equation, however, was not used and cannot be used for the argument of Aharonov et al.  \cite{pigeons2016}. The equation is true, but   not relevant.

  Let us reproduce the original argument for the PCP failure in a way that clarifies the inapplicability of (\ref{2}). Aharonov et al. make two claims:
  
  i) Given  pre- and postselection of three particles in two boxes in states  $|+++\rangle$ and $| +i+i+i\rangle$, a  single strong von Neumann measurement of any pair of particles  being together in the same box will show a null result with certainty, $\Pi_{ab}=0$. 
  
  ii) Given the above  pre- and postselection of three particles, the effect of weak bipartite interactions between all pairs of particles present in the same box disappear in the first order of the coupling strength.
 
  From (\ref{1}) and $\langle +i+i+i|+++\rangle\neq 0$,  we obtain the weak value \cite{AAV} of the projection operator corresponding to a pair of particles  being together in the same box:
  \begin{align}\label{wv}
 (\Pi_{ab})_w\equiv\frac{\langle +i+i+i|\Pi_{ab}|+++\rangle}{\langle +i+i+i|+++\rangle}=0;~~~~~{\rm for~all}~~a,b \in \{0,1,2\}.  
\end{align}
  Claims (i) and (ii) both follow from this equation. 

Aharonov and Vaidman \cite{AV91} proved a theorem stating that if the weak value of a  dichotomic variable is equal to an eigenvalue, then a strong measurement of this variable will yield this eigenvalue with certainty. $\Pi_{ab}$ is a dichotomic variable and 0 is one of its eigenvalues. Thus (i) follows  from (\ref{wv}).

The weak value of a projection operator characterizes the modification of all local weak interactions  \cite{PNAS}.  Therefore,  (ii) also follows  from (\ref{wv}). Note that the weak coupling of any particular pair does not  significantly  change the forward and backward evolving states, and thus $(\Pi_{ab})_w=0$ remains valid to first-order in the coupling strength for all pairs also when  weak interactions between all particles are present. 

 Kunstatter et al.   consider $\Pi$ instead of $\Pi_{ab}$. 
  From (\ref{2}) and $\langle +i+i+i|+++\rangle\neq 0$ we obtain
    \begin{align}\label{wvPi}
 (\Pi)_w\equiv\frac{\langle +i+i+i|\Pi|+++\rangle}{\langle +i+i+i|+++\rangle}=0.   
\end{align}
 $\Pi\equiv \Pi_{01}+ \Pi_{12}+ \Pi_{02}$ is a dichotomic variable, but its eigenvalues are 1 and 3 (see \cite{kunstatter2020escape} for derivation), so the weak value 0 is not equal to any of its eigenvalues. Thus we can learn nothing about the probability of outcomes of strong measurements from (\ref{wvPi}). Statement (i) does not follow from (\ref{wvPi}).

Statement (ii) also does not follow from (\ref{wvPi}). The claim in (ii) is about the separate effects of all the weak couplings between pairs of particles, even when they are applied in parallel. The weak value $(\Pi)_w =0$  only describes a single global property of these bipartite couplings. The weak value of a sum is the sum of the weak values, but the fact that the sum of weak values of projections vanishes tells us little about each individual weak value, since in general, it can be any complex number. 

 Kunstatter et al. considered operator identities 
 assuming the eigenvalue - eigenstate link. They showed, not surprisingly, that there is no  (eigen)state of an operator describing three particles in two boxes such that no two particles are together. To find such a situation would be a mathematical contradiction. The quantum violation of the  PCP \cite{pigeons2016,pigeonbook} is not more and not less than statements (i) and (ii). 
  Kunstatter et al., however,  read \cite{pigeons2016} in a different way. Representing Aharonov's proof they wrote in Section 1:
\begin{quote}
Thus they [Aharonov et al.] conclude that, for these particular pre- and post-selected elements of the ensemble, ``no two particles are in the same box'' during the transition {\it even when the} $\Pi_{ab}$ {\it operations are not applied}.
\end{quote}
Kunstatter et al. explicitly mentioned that the words in italic are their addition. Similarly, in Section 4 they wrote:
\begin{quote}
  Their  [Aharonov et al.'s] point is that measuring (i.e. projecting onto) the final state $|+i+i+i \rangle$ cancels all branches of the initial state in which $a$ and $b$ are in the same box. Due to the symmetry of the initial and final states, this is true for any choice of $~a,b \in \{0,1,2\} $. 
\end{quote}
These quotations show that Kunstatter et al. interpret the quantum violation of the PCP presented in \cite{pigeons2016} in classical terms: if it is true for any pair, it is true for all pairs together.  Aharonov et al., however, write: ``given the above pre- and post-selection, we have three particles in two boxes, yet no two particles can be found in the same box...'' 
This statement is about a measurement performed on a single pair.
Although  only one pair is measured, the statement is about all pairs because it is true for any pair.


The PCP is a principle of  classical logic, so it is not surprising that it cannot be violated with an eigenvalue - eigenstate link connecting the classical and quantum domains. The violation takes place for a pre- and postselected system for which, similarly to the Hardy paradox setup \cite{Hardy+,HardyAharonov},  ``elements of reality'' as certain results of strong measurements or as surprising weak traces do not correspond to eigenstates \cite{PRLmy}.


\section*{Acknowledgements}

This work has been supported in part by the National Science Foundation Grant   No. 1915015, U.S.-Israel Binational Science Foundation Grant No. 735/18, and the PBC PostDoctoral Fellowship at Tel Aviv University.
Y.A. was supported by grant number (FQXi-RFP-CPW-2006) from the Foundational Questions Institute and Fetzer Franklin Fund, a donor advised fund of Silicon Valley Community Foundation.

\bibliographystyle{unsrt}

\bibliography{bib_com}

\end{document}